\documentclass[runningheads]{llncs}
\usepackage[T1]{fontenc}
\usepackage{pdfpages}
\usepackage{times}
\usepackage{soul}
\usepackage{url}
\usepackage[hidelinks]{hyperref}
\usepackage[utf8]{inputenc}
\usepackage[small]{caption}
\usepackage{graphicx}
\usepackage{amsmath}
\usepackage{booktabs}
\usepackage{algorithm}
\usepackage{algorithmic}
\usepackage{paralist}
\usepackage{xspace}
\usepackage{xcolor}
\usepackage{units}
\newcommand{\mme}{MME\xspace}
\begin{document}
%
%\title{Contribution Title\thanks{Supported by organization x.}}

\title{Extracting Symbolic Models of Collective Behaviors
  with Graph Neural Networks\\
  and Macro-Micro Evolution}
 
\titlerunning{Extracting Symbolic Models of Collective Behaviors}
\author{%
Stephen Powers\orcidID{0000-0002-8918-8163}
% \and
% Joshua Smith\orcidID{0000-0001-7695-0027}
\and
Carlo Pinciroli\orcidID{0000-0002-2155-0445}}
%
%\authorrunning{Powers \emph{et al.}}
\authorrunning{Powers and Pinciroli}
\institute{Dept. of Robotics Engineering, Worcester Polytechnic Institute, Worcester, MA 01609, USA\\
\email{\{spowers2,cpinciroli\}@wpi.edu}}
\maketitle

\begin{abstract}
  Collective behaviors are typically hard to model. The scale of the swarm, the large number of interactions, and the richness and complexity of the behaviors are factors that make it difficult to distill a collective behavior into simple symbolic expressions. In this paper, we propose a novel approach to symbolic regression designed to facilitate such modeling. Using raw and post-processed data as an input, our approach produces viable symbolic expressions that closely model the target behavior. Our approach is composed of two phases. In the first, a graph neural network (GNN) is trained to extract an approximation of the target behavior. In the second phase, the GNN is used to produce data for a \emph{nested} evolutionary algorithm called \emph{macro-micro evolution (\mme)}. The macro layer of this algorithm selects candidate symbolic expressions, while the micro layer tunes its parameters. Experimental evaluation shows that our approach outperforms competing solutions for symbolic regression, making it possible to extract compact expressions for complex swarm behaviors.
\end{abstract}

\section{Introduction}
\label{sec:introduction}
Biological collective systems, such as fish schools and bird flocks, typically involve large numbers of individuals engaging in massively numerous interactions, with non-linear effects that produce complex and coordinated swarm-level behaviors. Identifying simple, yet effective, models to capture such interactions is typically difficult and it involves non-trivial data analysis and hypothesis testing \cite{camazineSelfOrganizationBiologicalSystems2003}.

In this paper, we propose an automatic approach to symbolic regression designed to facilitate modeling of collective behaviors. Given the raw data of the collective behavior, e.g., the trajectories of the agents or information about their pairwise interactions, our approach derives a human-readable symbolic expression that best approximates the input data. Our work is applicable to both natural sciences (e.g., to discover new models of collective behaviors), and in engineering (e.g., to extract approximate models of collective behaviors optimized in a centralized manner).

Our method is based on techniques in machine learning and evolutionary computation. These techniques have been applied extensively for the design of artificial collective systems such as swarms of robots. A large body of work exists that uses neural networks trained through evolutionary algorithms (e.g., genetic algorithms) to design specific collective behaviors \cite{trianniEngineeringEvolutionSelfOrganizing2011,trianniEvolutionSelforganizationSwarm2008}. The same techniques have also found applications in fault detection among robot swarms \cite{christensenFaultDetectionAutonomous2008}. General-purpose methods for system identification similar to generative adversarial networks have also been proposed, e.g. Turing Learning \cite{liTuringLearningMetricfree2016}. While effective in capturing the desired collective behaviors, the neural networks produced by these models are typically black boxes not amenable to analysis, hypothesis testing, and correction.

For this reason, several works aim to produce interpretable behavioral models obtained through optimization and evolutionary methods. Notable examples include the variants of Automode \cite{birattariAutoMoDeModularApproach2021,francescaAutoMoDeChocolateAutomaticDesign2015}, novelty search to discover collective behaviors in minimalistic robots \cite{brownDiscoveryExplorationNovel2017}, grammatical evolution \cite{ferranteGESwarmGrammaticalEvolution2013}, and behavioral trees \cite{neupaneLearningSwarmBehaviors2019,kucklingBehaviorTreesControl2018}. All these works share the objective of producing a desired (and unknown) behavior for artificial agents (robots), rather than extracting interpretable models from existing data.

To the best of our knowledge, no work has been explicitly devoted to symbolic regression for collective behaviors. The main contribution of our paper is therefore to propose the first method to perform this task. Our approach is composed of two phases:
\begin{enumerate}
\item In the first phase, the input data, both raw and post-processed, is used to train a graph neural network (GNN).
\item In the second phase, a symbolic expression is extracted using a nested genetic algorithm we refer to as \emph{macro-micro evolution} (\mme). The fitness of the symbolic expression is calculated using data generated by the GNN.
\end{enumerate}
The power of our method comes from the combination of these two phases. GNNs offer embeddings that are ideal to capture the many facets of collective phenomena, such as pairwise interactions among agents, as well as diffusive and aggregative processes across agents. The \mme decomposes the problem of system identification into two nested problems:
\begin{inparaenum}[\it (i)]
\item identification of the structure of a parametric expression; and
\item estimation of the parameter values.
\end{inparaenum}

For validation, we show that our approach can capture three compelling variants of collective behaviors:
\begin{inparaenum}[\it (i)]
\item Hexagonal shape formation among homogeneous agents;
\item Square shape formation among heterogeneous agents;
\item Coordinated motion behaviors (i.e., boids \cite{reynoldsFlocksHerdsSchools1987}), which combine pairwise interactions and data aggregation.
\end{inparaenum}

We compare our approach with several state-of-the-art methods for symbolic regression. Our analysis reveals that, while hexagonal shape formation can be solved by most methods, our method solves it more efficiently. In addition, our method is capable to attack problems, such as square shape formation and coordinated motion, which are not solvable by existing methods.

Our paper is organized as follows: in Sec. \ref{sec:related_work} we review existing work on symbolic regression and highlight the novelty of our approach. In Sec. \ref{sec:methodology} we present our approach. In Sec. \ref{sec:experiments} we report the results of our analysis. We conclude the paper in Sec. \ref{sec:conclusion}.

\section{Related Work}
\label{sec:related_work}

The problem of symbolic regression is ubiquitous in science and engineering. Given a data set as input, the objective is to produce a symbolic expression (i.e., a mathematical formula) that closely fits the data. The constant growth, in both size and number, of available datasets makes it increasingly more desirable to automate this process. Depending on the nature and size of the dataset, the search for suitable symbolic expressions can be extremely hard. With complex, non-linear phenomena, the search space of possible expressions may suffer from combinatorial explosion. 

A common approach to automate symbolic regression involves genetic algorithms \cite{19:Orze}. An expression is represented as a syntax tree formed by mathematical operators, fundamental functions, and operands. The genetic algorithm modifies the expressions through mutation and crossover. At each iteration, the expressions are ranked according to metrics such as accuracy vs. the input data and complexity of the expression \cite{17:Chen,15:Huang,16:Motta,20:White,18:Zhong}. A prominent approach is Epsilon-Lexicase Selection \cite{Eplex}.

AI Feynman \cite{21:Sil} is an alternative approach that tames combinatorial explosion through a diverse suite of techniques. These techniques include providing the search algorithm with a database of known equations from physics, performing dimensional analysis, attempting polynomial fits, training a neural network as a data interpolator, and transform the data in various ways. Extensive tests performed on well-known physics equations yielded near-perfect performance, surpassing Epsilon-Lexicase Selection. However, our analysis in Sec. \ref{sec:experiments} shows that the performance of AI Feynman on datasets of collective behaviors is far from satisfactory.

% These algorithms provide promising results on standard regression analysis problems in which one must fit a symbolic model to raw data. However, these algorithms are unable to extract symbolic models representing a variable number of individuals in a group. This is due to the fact that these algorithms assume a constant number of variables to consider. For example, running these algorithms on data from a group of 10 individuals will produce a drastically different model than running the algorithms on data from a group of 100 individuals.

Neural networks \cite{Kaufmann,Qingbiao,Ried,tolstaya,Ward_Gobet_Kendall_2001} are another approach to model collective behaviors. In particular, Cranmer \emph{et al.} proposed a method to perform symbolic regression in particle systems that utilizes a GNN as a data interpolator rather than the final product \cite{Cranmer}. Similarly to our approach, the output of the GNN is fed to an evolutionary algorithm designed for symbolic regression. Cranmer \emph{et al.} employed Eureqa — a commercial suite of machine learning algorithms — and a self-developed Python-based library (PySR) \cite{PSR}. Both Eureqa and PySR attempt to evolve the expression tree along with its parameters in a single step. We build upon this work to show that a \emph{nested}, \emph{hierarchical} genetic algorithm greatly enhances the performance of symbolic regression.

\begin{figure}[t]
  \centering
  \includegraphics[width=1\linewidth, angle=0]{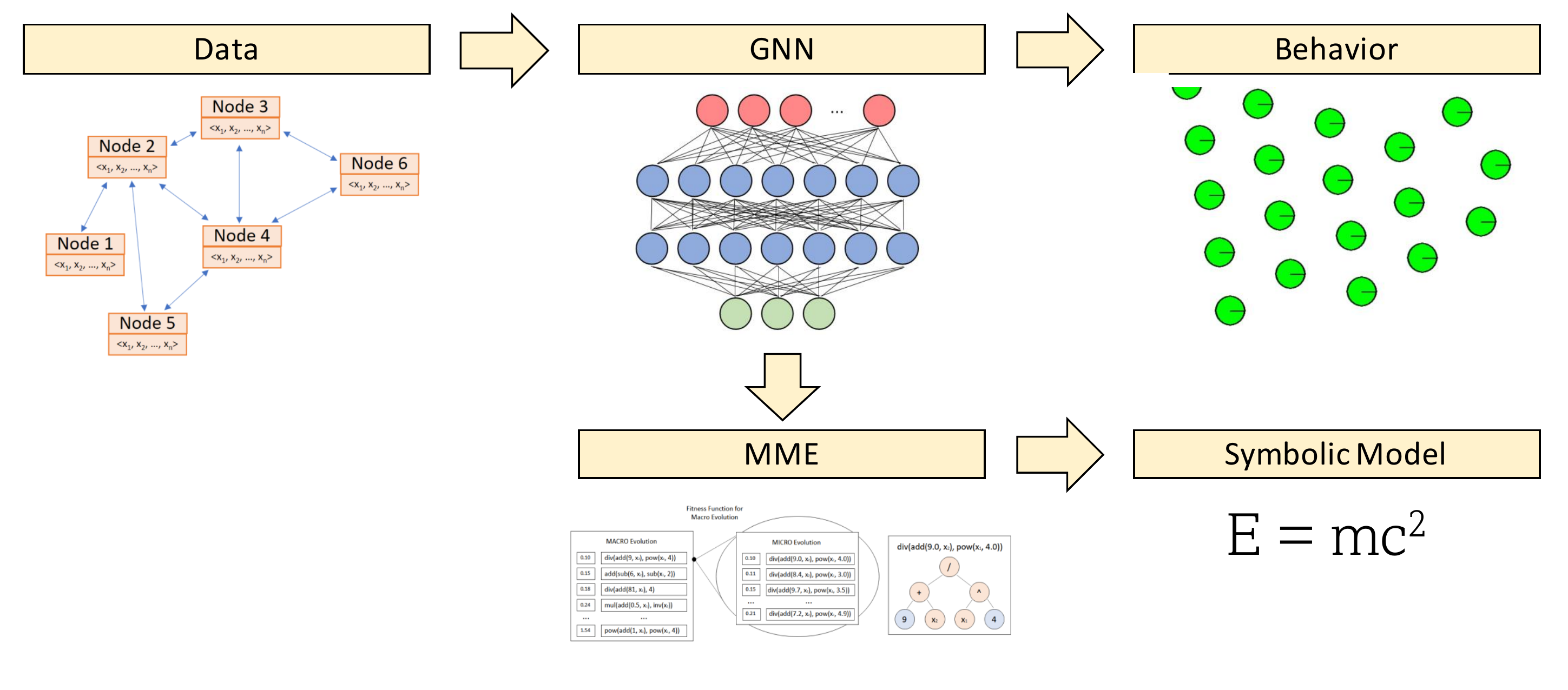}
  \caption{Overview of our approach.}
  \label{fig:overview}
\end{figure}

\section{Methodology}
\label{sec:methodology}

\subsection{System Overview}
\label{sec:overview}

As mentioned, our approach is comprised of two phases. In the first, we train a GNN to model the target collective behavior. In the second, we employ a nested genetic algorithm, \mme, to select appropriate symbolic expressions. The macro layer selects a parametric expression, while the micro layer tunes the expression’s parameters. Data generated by the GNN is used in the \mme to evaluate the fitness of each candidate expression.

\subsection{Input: Data and Priors}
Our approach accepts raw data as input. For example, in modeling collective motion, one could feed our approach with the poses of the individual agents. However, it is often possible to hypothesize that certain post-processed expressions of the raw data might be handy, e.g., distance between agents or its inverse. We call these expressions \emph{priors}. Including priors eliminates the need to ``rediscover'' them, making symbolic regression faster and more accurate. We highlight that, in practice, there is no harm in providing priors that are unrelated to the final expression, because both phases are able to ignore the unnecessary inputs. This makes it possible to create ``libraries'' of common priors taken from the literature, taking advantage of our approach to identify the correct priors. This creates an interesting feedback loop for researchers engaged in model identification, as the type of inputs selected for the final expression offer insight on the nature of the mechanisms involved in the target collective behavior.

% We can optimize the learning of the GNN and MME by preprocessing the raw data into primitive values that we --- as the end user --- suspect will aid in understanding the relationships between individuals in the swarm. For instance, in just about all spatially-dependent collective behaviors, the distance between individuals is critical in determining how the individuals will interact. As such, we are able to eliminate the need for the neural network to learn an approximation of the euclidean distance formula as well as the interactions by calculating the distance prior to submitting the data for the GNN to learn. This preprocessing allows us to combine our expert knowledge of general swarm behaviors with the power of the machine learning. 

\subsection{Phase 1: Graph Neural Networks}
\label{sec:gnn}
Graphs are natural models of swarm behaviors. Each node typically represents an agent and edges capture pairwise interactions and relationships. GNNs are neural networks superimposed on a graph structure. In a GNN, nodes, edges, and even the entire graph are associated with neural networks, respectively referred to as the \emph{node model} ($\phi^n$), the \emph{edge model} or \emph{message function} ($\phi^e$), and the \emph{graph model} ($\phi^g$). The node model captures the individual behavior of an agent $i$ as a result of internal state and pairwise interactions with neighbors, e.g., the sum of virtual forces in hexagonal formation. The output of the node model is the \emph{node state} $x_i$. The edge model represents the interaction between two agents $i$ and $j$, e.g., the Lennard-Jones potential \cite{smitPhaseDiagramsLennard1992}. Its output is called a \emph{message} $y_{ij}$. The graph model aggregates the states $x_i$ and messages $y_{ij}$ to yield a swarm-level representation $z$, e.g., the regularity of the hexagonal pattern. These models are the ultimate targets of symbolic regression, depending on the application at hand. From a mathematical standpoint, a GNN can be formalized as follows:
\begin{align*}
  x_i[t+1] &= \phi^n(x_i[t], \sum_{j \in \mathcal{N}_i} y_{ij}[t]) \\
  y_{ij}[t+1] &= \phi^e(x_i[t], x_j[t]) \\
  z[t+1] &= \phi^g(x_i[t+1], \dots, y_{ij}[t+1], \dots)
\end{align*}
where $\mathcal{N}_i$ indicates the neighborhood of agent $i$ and $t$ is the iteration index.

% \begin{figure}[t]
%   \centering
%   \includegraphics[angle=270, width=0.85\linewidth]{GNNOverview.pdf}
%   \caption{Comparison of GNN structure to collective behaviors.}
%   \label{fig:gnnoverview}
% \end{figure}

\subsection{Phase 2: Symbolic Modeling with Macro-Micro Evolution}
\label{sec:mmevo}

\subsubsection{Fitness function.}
\mme is a nested evolutionary algorithm in which the outer (macro) evolution selects the structure of an expression $\lambda$, and the inner (micro) evolution tunes its parameters. Both evolutions use the same fitness function $f(\lambda)$, which linearly combines two functions $f^c(\cdot)$ and $f^a(\cdot)$ of the \emph{complexity} and the \emph{accuracy} of $\lambda$ according to a weight $\rho \in [0,1]$:
\begin{equation}
  \label{eq:scoring}
  f(\lambda) = \rho \cdot f^c(\text{complexity}(\lambda)) + (1 - \rho) \cdot f^a(\text{accuracy}(\lambda)).
\end{equation}

\subsubsection{Complexity.}
To measure the complexity of an expression $\lambda$, we first associate a cost to each operator. For example, additions and subtractions, being simple operators, could be assigned a cost of 2; exponentiation of the form $2^{\lambda}$ could be assigned a cost of 20. The complexity of an expression is then calculated as the sum of the costs of each operator that appears in it. The user can define what operators should be considered, along with the costs to using them. As a parameter to \mme, the user can also assign a target complexity $\tau$ for the final expression, which indicates the acceptable complexity of the final expression. Tuning $\tau$ is a way to prevent overfitting. The function $f^c(\cdot)$ that transforms the complexity is defined as
\begin{equation*}
f^c(\lambda) = \frac{\text{max}(0, \text{complexity}(\lambda) - \tau)}{\tau}.
\end{equation*}

\subsubsection{Accuracy.}
We determine an individual's accuracy via the Mean Square Error (MSE) between the answers generated by an expression $\lambda$ and the data generated by the GNN. Any expression $\lambda$ that, once evaluated, results in an undefined or infinite value, or a complex number, is discarded. To prevent large values of MSE from rendering the contribution of complexity negligible, we define
\begin{equation*}
h(\lambda) = \frac{\text{MSE}(\lambda)}{\text{MSE}(\lambda^{\text{worst}})}
\end{equation*}
where $\lambda^{\text{worst}}$ represents the least accurate expression within the subset of surviving expressions of the generation.

\begin{figure}[t]
  \centering
  \includegraphics[width=1\linewidth, angle=0]{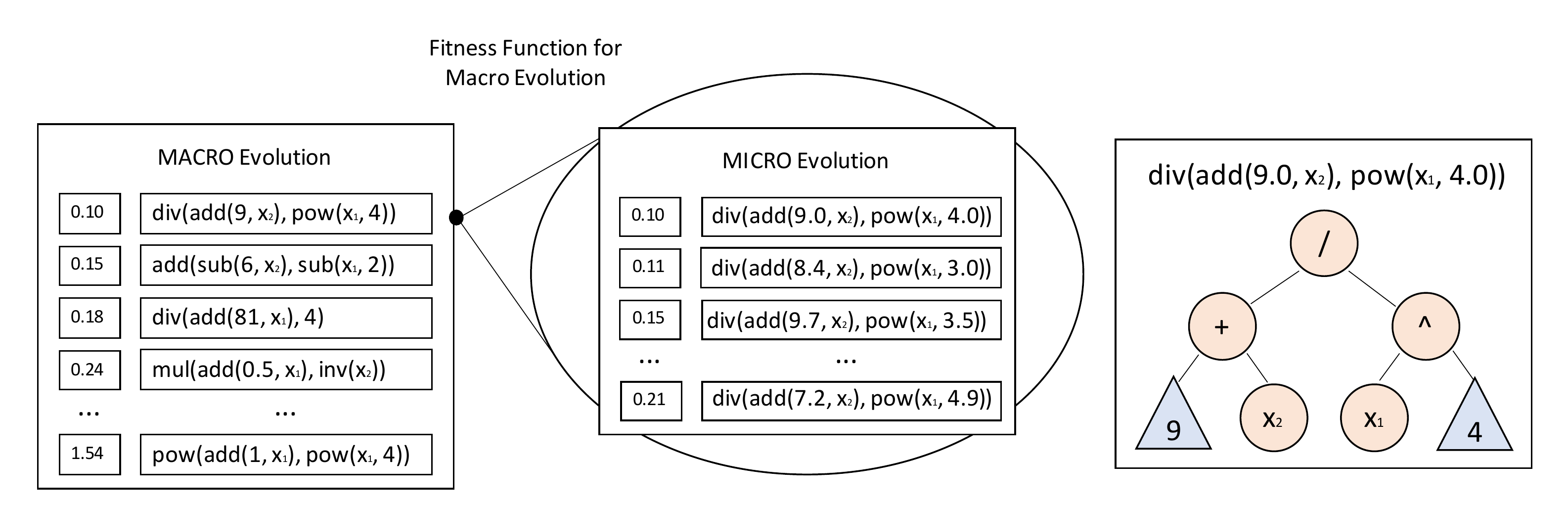}
  \caption{Representation of the difference between macro and micro phases of MME. Circular nodes (structural) are modified in the macro phase while triangular nodes (parameters) are modified in the micro phase. }
  \label{fig:MMEExplanation}
\end{figure}

\subsubsection{Micro-macro evolution.}
At the start of each generation, \mme performs \emph{selection} and \emph{duplicate removal}. Selection involves picking a subset of high-fitness expressions (the ``parents'') and the creation of ``children'' through crossover and mutation. Two individuals are considered duplicates if they share the same structure, regardless of the value of their parameters. Two passes of scoring then occur. The first pass ranks the new children and the surviving parents. Those that are bound to survive to the next generation are then run through the second pass, i.e., the micro-evolutionary algorithm, to determine the optimal parameter values as shown in Figure \ref{fig:MMEExplanation}. While this could theoretically result in long run times, in practice the micro evolution reaches convergence in a small number of generations.

\section{Experimental Evaluation}
\label{sec:experiments}
To validate the effectiveness of our approach, we considered three case studies with complementary features: hexagonal shape formation, square shape formation, and coordinated motion. A video that showcases the resulting collective behaviors found by our approach against the original models is available at \url{https://youtu.be/r6r5GBH7Iuk}.

\subsubsection{Hexagonal shape formation.}
Shape formation is a well-studied problem in both natural and artificial swarms. The typical model that achieves hexagonal shapes imagines a swarm of identical agents immersed in an isotropic virtual potential field, in which the distance between pairs of agents induces an interaction force between them. The Lennard-Jones (LJ) potential \cite{smitPhaseDiagramsLennard1992} is one of the most common models of interaction due to its simplicity:
\begin{equation*}
  V_\text{LJ}(r; \delta, \epsilon) = 4 \epsilon \left( \left(\frac{\delta}{r}\right)^{12} - \left(\frac{\delta}{r}\right)^6 \right)
\end{equation*}
where $r$ is the current distance between the agents, $\sqrt[6]{2}\,\delta$ is the distance at which the potential is minimum, and $\epsilon$ is the depth of the minimum. Differently from the electric and gravity potentials, the LJ potential has both an attractive and a repulsive component. The LJ potential is a good testbed to verify that the edge model $\phi^e$ of a GNN can correctly capture the interaction force between identical individuals, and that \mme can derive a satisfactory approximation of the virtual interaction force $F_{\text{LJ}} = -\partial V_{\text{LJ}}(r; \delta, \epsilon) / \partial r$.

\subsubsection{Square shape formation.}
A straightforward extension of hexagonal formation is square formation. We divide the agents into two categories, e.g., by color, and use different LJ potentials depending on the color of the agents. If two interacting agents belong to different categories (i.e., have different colors, referred to as \emph{non-kin}), their target distance is $\delta$; if the agents belong to the same category (\emph{kin}), their target distance is $\sqrt{2}\,\delta$. Square formation tests whether the GNN can correctly classify the two types of interaction in its edge model $\phi^e$, and whether symbolic regression is able to construct appropriate expressions.

\subsubsection{Coordinated motion.}
Coordinated motion (flocking) is a compelling test case for our approach, due to the complex symbolic form of the control law followed by the agents. We use the classical boids model by Reynolds \cite{reynoldsFlocksHerdsSchools1987}, in which the speed of an agent is calculated as the weighted sum of three components that account for separation, alignment, and cohesion. For an agent, \emph{separation} is defined as the average of the repulsion forces to neighbors that are closer than a certain threshold. \emph{Alignment} is defined as the average of the velocities ($\vec{v}_j$) of the neighbors $j$ of an agent. \emph{Cohesion} is a unit vector that points to the centroid of the position of the neighbors. Combination of these three terms results in Eq. \ref{eq:boids} where $C$, $S$, and $A$ are user-defined weighted values of the cohesion, separation, and alignment terms respectively, $|\mathcal{N}|$ is the number of neighbors, $\vec{x}_j$ is the position of a neighbor within the boid's local coordinate frame, and $||\vec{x}_j||$ its length:
\begin{equation*}
  \label{eq:boids}
  F_{\text{boids}}(\vec{x}_j, \dots, \vec{v}_j, \dots) =
  \frac{1}{|\mathcal{N}|}
  \sum_{j \in \mathcal{N}}\left(
    \text{C}\frac{\vec{x}_j}{||\vec{x}_j||} -
    \text{S}\frac{\vec{x}_j}{||\vec{x}_j||^2} + 
    \text{A}\vec{v}_j
  \right).
\end{equation*}
Coordinated motion allows us to verify the ability of a GNN to capture pair-wise inter-agent interaction in the edge model $\phi^e$, as well as the aggregation of such interaction in the node model $\phi^n$.

\subsubsection{Data generation.}
We followed a similar methodology across the three case studies. In all of them, we ran simulations to generate a sufficient amount of data. The agents in were represented as holonomic point-masses operating in two dimensions with coordinates ranging between 0 and \unit[1]{m}. Motion wrapped around the environment boundaries, making it a torus. For simplicity, the mass of each robot was set to \unit[1]{kg}, and we used a double integrator to determine dynamics. The actuation was the acceleration vector of each robot. Sensing and communication were assumed noiseless, and their range was \unit[0.5]{m}. The $\delta$ parameter for hexagonal shape formation was set to \unit[0.13]{m}. For square shape formation, the parameters were $\delta_{\text{kin}} = \unit[0.13]{m}$ and $\delta_{\text{non-kin}} = \sqrt{2}\,\delta_{\text{kin}}$. As for boids, we set $C = 2$, $S = 75$, and $A = 3$. We used 20 agents for the shape formation experiments and 50 for boids. Overall, for each setup, we ran 250 simulations, each 25 simulated seconds long. We recorded data at \unit[10]{Hz} for shape formation experiments and \unit[30]{Hz} for boids. After the simulations, we post-processed positional and angular data to be between 0 and 1 for all inputs. All data passed to the GNN was represented in Cartesian coordinates.

\subsection{Accuracy of GNN Models}
\label{sec:gnn_accuracy}

\begin{figure}[t]
  \centering
  \includegraphics[width=\linewidth]{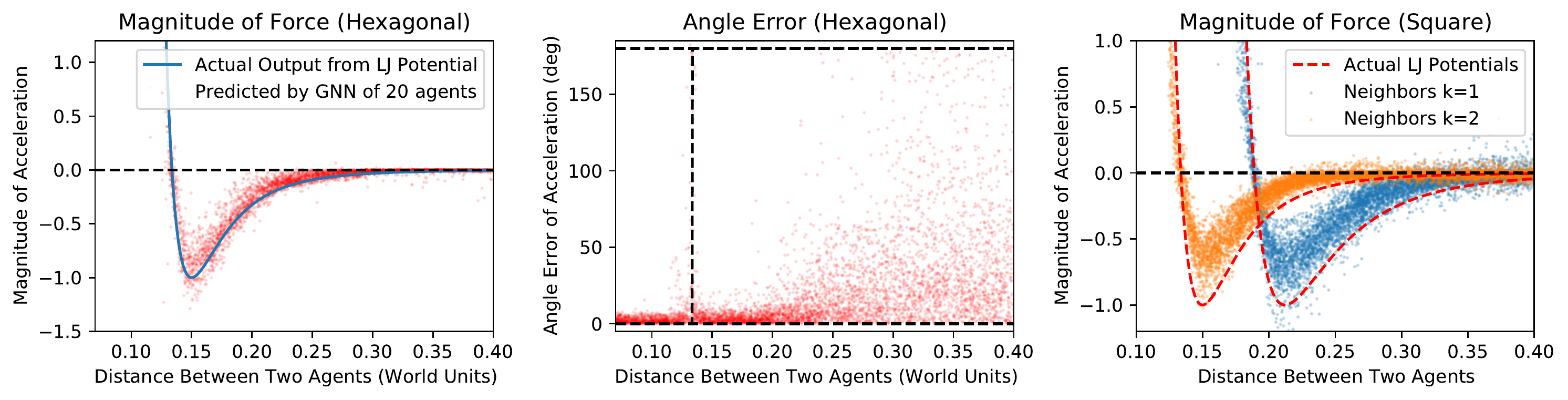}
  \caption{The outputs of the GNN relative to the actual LJ potentials for the hexagonal (left) and square formations (right), as well as the errors of the angle of the output vector of the GNN in the hexagonal formation (middle)}
  \label{fig:hexErr}
\end{figure}

\subsubsection{GNN configuration.} We employed the same type of GNN in all our experiments. The GNN used 4 fully connected layers: an input and an output layer whose sizes depend on the case study, and two hidden layers with 300 neurons each.

\subsubsection{Hexagonal Lattice.}
We trained a GNN for 200 epochs with the Adam optimizer. To validate its performance, we ran the trained GNN with two inputs. The first input represents an agent at the origin of the world, and the second represents an agent at a random distance from the first. The output from the GNN represents the virtual force on the first agent expressed in Cartesian coordinates. Fig. \ref{fig:hexErr} (left) compares the magnitude of the force vector due to the original LJ potential used in simulation with 4,500 random outputs from the GNN, showing remarkable similarity between the two. Fig. \ref{fig:hexErr} (middle) reports the error in the angle of the force vector between the GNN and the reference vector. The error is next to zero in the range $[0,0.2]\unit{m}$, which corresponds to the area where the interaction between agents is most significant. The error increases with distance beyond this range.

\subsubsection{Square Lattice.}
We established the same setup for the square lattice case. We obtained the same amount of data as was obtained in the hexagonal lattice experiment, with the addition of edge attributes. Connections between kin neighbors were given an attribute of 1, while non-kin had an attribute of 2. We generated 10,000 pairs of nodes separated by random distances (between $\unit[0.07]{m}$ and $\unit[0.4]{m}$) and provided them as input to the GNN. 5,000 data points contained edge attributes of 1, and 5,000 data points contained an edge attribute of 2. The magnitude of the forces generated by the GNN compared to the reference are reported in Fig. \ref{fig:hexErr} (right). The GNN correctly categorized the kin and non-kin interactions, showing acceptable accuracy in capturing both potentials.

\subsubsection{Boids.}
For the boids experiment, we initially attempted to train the GNN using solely raw data, but observed degraded performance. We then decided to provide the network with priors. These were distance and velocity vectors of a neighbor relative to the individual of interest, as well as their magnitudes and the inverse of these magnitudes. We also included a normalized version of these vectors. An example of the resulting behavior compared to the reference can be seen in the video at \url{https://youtu.be/r6r5GBH7Iuk}.

\subsection{Symbolic Models}
\label{sec:sym_models}

\begin{table}[t]
  \caption{Symbolic models found with the considered approaches.}
  \label{tab:finalEquations}
  \scriptsize

  \begin{tabular}{ccccc} \toprule
    \hline
    \addlinespace
    Case Study & Target Equation & Algorithm & Resulting Equation\\
    \addlinespace
    \hline

    \addlinespace
    Hex & \(\left(\dfrac{1.2e-10}{x^{12}}\right) - \left(\dfrac{2.2e-5}{x^{6}}\right)\) & \mme & \( \left(\dfrac{8e-9}{x^{10.07}}\right) - \left(\dfrac{9.8e-6}{x^{6.54}}\right)\)  \\ 

    \addlinespace
    &  & PySR & \(\dfrac{-\left(0.42-\dfrac{0.06}{x}\right)}{\left(x-0.02\right)\left(8.14e8x^{11.66}-x+0.272\right)} - 0.04\)  \\ 
    
    \addlinespace
    &  & EpLex & \(2.38x - 1.3 + \left(\dfrac{0.15}{x}\right) \)   \\ 
    
    \addlinespace
    & & AIF & \( \left(5.6e-5\left(x+\dfrac{x^{0.5}}{-\left(x+2\right)} \right) \right)^{0.5}  \)   \\ 
    
    \addlinespace
    \hline

    \addlinespace
    Square (Kin) & \( \left(\dfrac{7.84e-9}{x^{12}}\right) - \left(\dfrac{1.7e-4}{x^{6}}\right)\) & \mme & \( \left(\dfrac{6.8e-7}{x^{9.75}}\right) - \left(\dfrac{1.9e-5}{x^{7.75}}\right)\)    \\

    \addlinespace
    Square (NonKin) & \( \left(\dfrac{1.2e-10}{x^{12}}\right) - \left(\dfrac{2.2e-5}{x^{6}}\right)\) & \mme & \( \left(\dfrac{1.58e-9}{x^{10.67}}\right) - \left(\dfrac{4e-6}{x^{6.79}}\right)\)  \\

    \addlinespace
    \hline

    \addlinespace
    Boids & \( 2\dfrac{\vec{x}}{||\vec{x}||}  - \left(\dfrac{75}{||\vec{x}||}\right)\dfrac{\vec{x}}{||\vec{x}||} + 3\dot{\vec{x}} \) & \mme & \( 0.59\dfrac{\vec{x}}{||\vec{x}||} - \left(\dfrac{1}{||\vec{x}||}\right)\dfrac{\vec{x}}{||\vec{x}||} + \left(\dfrac{1}{||\dot{\vec{x}}||}\right) \dot{\vec{x}}\)
    
    \tabularnewline
    \addlinespace
    \hline

    \bottomrule
    
  \end{tabular}
\end{table}

%\subsubsection{Algorithm Comparability}
We compared \mme to three other state-of-the-art symbolic regression algorithms: AIFeynman (AIF) \cite{21:Sil}, Epsilon-Lexicase (EpLex) \cite{Eplex}, and PySR \cite{PSR}. For all three evolutionary algorithms (\mme, PySR, and EpLex), we used a population size of 4,000, and limited the operators to add, subtract, multiply, divide, and power. Each run was allowed a maximum of 200 generations. We capped to 3 and 15 the minimum and maximum number of operands in an expression for EpLex. AIFeynman was set up differently due to its capabilities. The algorithm---to the best of our knowledge---does not consider power operators; as such, the only operators given to it were add, subtract, multiply, divide, negate, and invert. We specified the maximum number of polynomial terms in a given expression to 6, and allowed the interpolating neural network to train up to 200 epochs.

\subsubsection{Hexagonal Lattice.}
To test the effectiveness of using symbolic regression to reverse engineer the data generated by a GNN, we reused the 4,500 points shown in Fig. \ref{fig:hexErr} (left) as input to all four algorithms. Each algorithm was run 30 times on the same data. We set a target complexity $\tau$ of 12 for \mme, where the average cost of operations used was equal to the arity of the operation, with the exception of the power operator in which the exponent was an operation instead of a constant being 20 instead of 2.

\begin{figure}[t]
  \centering
  \includegraphics[width=1\linewidth]{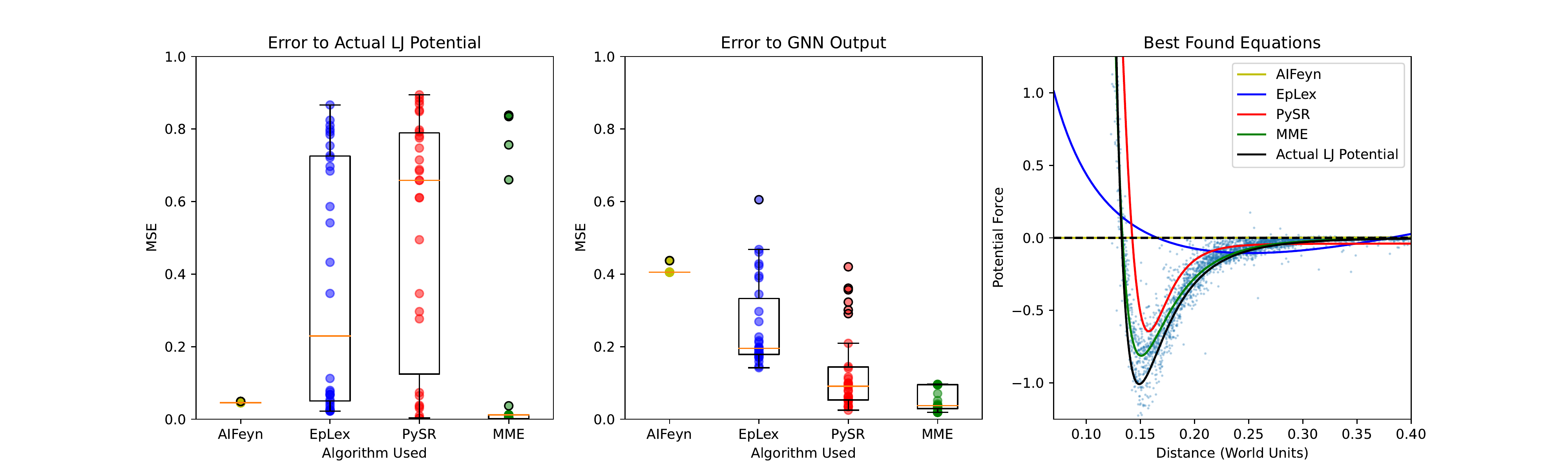}
  \caption{The error from 30 runs of each evolutionary algorithm related to the actual Lennard-Jones expression (left) and the GNN-generated output (middle) as well as the best output equation from each algorithm (right).}
  \label{fig:comparison}
\end{figure}

We calculated the error using the MSE in which all values were clipped between -1 and 1. These specific values allowed for easier analysis. PySR and EpLex created several solutions containing asymptotic sections within the domain of interest. This yielded very high errors in these sections, which skewed the results significantly. The errors from the 30 experiments are shown in Fig. \ref{fig:comparison} in relation to the actual LJ potential used in simulation and also to the GNN-generated output data. The best expression plotted in Fig. \ref{fig:comparison} is that which has the lowest MSE relative to the GNN output data. The best expression found within the 30 runs for each of the four algorithms is also shown in Tab.~\ref{tab:finalEquations}. It should be noted that AIFeynman resulted in low MSE solely due to the fact that complex numbers were ignored in this analysis.

As seen in Tab.~\ref{tab:finalEquations}, the expressions generated by \mme are remarkably similar to the target equation in structure. The only difference between the resulting expression and the target expression are the parameters, which, in a practical scenario, may be manually tuned as needed. No other state-of-the-art algorithm finds any expression remotely similar in structure to the target equation. Even if the found expressions are technically accurate in terms of MSE, they are unrecognizable when compared to the original reference.
 
\subsubsection{Square Lattice.}
Considering the similarities between the square and hexagonal shape formations, only \mme was tasked with finding the resulting equations shown in Tab.~\ref{tab:finalEquations}. Once again, \mme finds symbolic expressions identical in structure to the target equations. This case, however, provides an interesting observation into the resulting expressions parameters. While the parameters between the resulting expressions and the target expressions are different, in each case, the ratio between kin and non-kin are very similar. This results in a behavior that is indistinguishable from the original as seen in the video at \url{https://youtu.be/r6r5GBH7Iuk}.

\subsubsection{Boids.}
We used \mme to find a symbolic expression from the data from the GNN and the 12 priors described in Sec. \ref{sec:gnn_accuracy}. After running \mme 10 times on this data, we sorted solutions by complexity first and then by MSE. The top equation was simplified manually and can be found in Tab.~\ref{tab:finalEquations}. While not identical to the target equation, the resulting expression shows a strong similarity---each of the three terms (cohesion, separation, and alignment) can be clearly seen in the final expression. Overall, this shows the efficacy of adding priors combined with the architecture of our system. The video at \url{https://youtu.be/r6r5GBH7Iuk} shows that the final expression does not fully reproduce the flocking behavior of boids. This issue could be solved through manual tuning.

\section{Conclusion}
\label{sec:conclusion}
We presented an automatic approach to extract symbolic models of collective behaviors. Our approach combines the power of GNNs in approximating unknown network functions from data, and the versatility of grammatical evolution in manipulating symbolic expressions. We showed that, by employing a nested genetic algorithm called \mme, we can reproduce expressions that closely match the original expressions used as a reference. We validated the effectiveness of our approach using three well-known, but non-trivial, case studies: hexagonal pattern formation, square pattern formation, and collective motion. Our approach works well where other state-of-the-art symbolic regression algorithms may struggle.

Our approach does not remove the need for expert knowledge and ingenuity in deriving models of collective behaviors from real data. Hypothesis testing and experiment design remain in the hands of the researchers. However, we believe that automatically producing viable candidate expressions greatly facilitates the job, in at least two important ways. First, because the expressions we can produce are \emph{compact}: our approach explicitly promotes short expressions, which are human-readable and easier to analyze. Second, because our approach offers the means to understand which parts of the input (i.e., raw data or priors with a generic set of functions) matter. This second aspect is important, as it offers an informative form of feedback to drive better hypothesis testing and experiment design.

Future work includes replacing the mean square error with alternate methods of determining the accuracy of a signal, such as dynamic time warping and cross correlation coefficients. Other machine learning techniques such as Turing Learning and generative adversarial networks may also be used. The concept of dual-phase algorithms for symbolic regression may also be applied to other state-of-the-art algorithms, such as using ant colony optimization for establishing structure and particle swarm optimization for tuning parameters.

\section*{Acknowledgments}
This work was funded by a DCRG grant from MathWorks, Inc. Results in this paper were obtained in part using a high-performance computing system acquired through NSF MRI grant DMS-1337943 to WPI. The authors wish to thank Joshua Smith for his precious technical contribution to the realization of this work.

\bibliographystyle{splncs04}
\bibliography{carlo,stephen}
 
\end{document}